# Visualizing the robustness of query execution

Goetz Graefe, Harumi Kuno, Janet L. Wiener

Hewlett-Packard Laboratories, Palo Alto, CA

## Abstract

In database query processing, actual run-time conditions (e.g., actual selectivities and actual available memory) very often differ from compile-time expectations of run-time conditions (e.g., estimated predicate selectivities and anticipated memory availability). Robustness of query processing can be defined as the ability to handle unexpected conditions. Robustness of query execution, specifically, can be defined as the ability to process a specific plan efficiently in an unexpected condition. We focus on query execution (runtime), ignoring query optimization (compile-time), in order to complement existing research and to explore untapped potential for improved robustness in database query processing.

One of our initial steps has been to devise diagrams or maps that show how well plans perform in the face of varying run-time conditions and how gracefully a system's query architecture, operators, and their implementation degrade in the face of adverse conditions. In this paper, we show several kinds of diagrams with data from three real systems and report on what we have learned both about these visualization techniques and about the three database systems.

## 1  Introduction

If database query performance is consistently too slow (by a moderate factor), additional hardware and well-studied techniques for concurrency and for parallel query execution solve the problem at relatively low cost. If query performance is unpredictable, however, well-educated and well-paid database administrators must race to isolate, understand, and address each individual problem whenever it occurs. The usual sources are unexpected run-time conditions, e.g., errors in cardinality estimation or resource contention. We understand robustness of query processing as the ability to handle such unexpected conditions. Due to the expense involved, robustness of query performance is now as important as traditional performance techniques.

Much existing research into robustness focuses on poor plan choices during query optimization. Some techniques focus on cardinality estimation, run-time feedback to query optimization, dynamic re-optimization invoked during run-time, etc. [BBD05, BBDW05, CNR08, ML02, MRSLPC04, SLMK01, ZHJLZ05]. Others explore sophisticated management of query execution plans, e.g., caching multiple plans for a single query or automatic testing of new query execution plans [INSS98, ZDSZY08]. In contrast and as a complement to those efforts, we focus on the role of query execution techniques. We define robust query execution as the ability to execute a specific query execution plan efficiently under any conditions. I.e., we assume that query optimization is complete and the chosen query execution plan is fixed. We complement other efforts that assume that the query execution engine is unchangeable and all adaptive or robust techniques must reside in the query optimizer.

Our goal is to enable the measurement and comparison of how gracefully a database system's query architecture, operators, and their implementation degrade during adverse conditions. To that end, we provide several kinds of diagrams that we call robustness maps. These maps quantify and visualize how performance degrades as work increases or as resources decrease. Robustness maps permit reasoning about the executor's impact on query robustness. E.g., they can inform regression testing as well as motivate, track, and protect improvements in query execution.

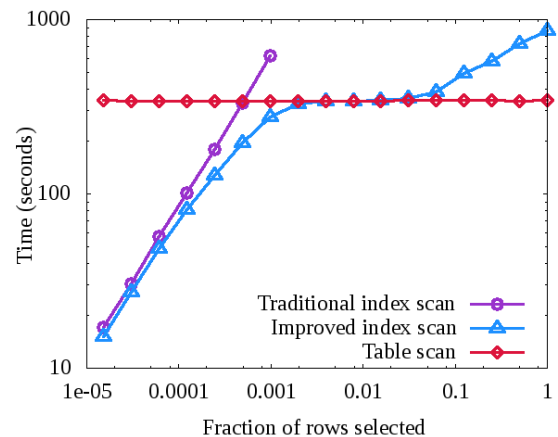

Figure 1. Single-table single-predicate selection.

Consider, for example, the diagram in Figure 1. It shows execution times for selecting rows from a table

(TPC-H line items, about 60M rows) for a variety of selectivities (result sizes). Selectivities and execution times both use logarithmic scales. Query result sizes differ by a factor of 2 between data points.

The performance of three query execution plans is shown. One plan is a traditional table scan – its performance is constant across the entire range of selectivities but for small result sizes it is unacceptably slow compared to the index scans.

The traditional index scan, on the other hand, is unacceptably slow for moderate and large result sizes due to the need to fetch qualifying rows from the table – its cost is so high that it is not even shown across the entire range of selectivities. The break-even point between table scan and traditional index scan is at about 30K result rows or $2^{-11}$ of the rows in the table.

The improved index scan, however, combines low latency for small results as well as high bandwidth for moderate result sizes. The cost of the improved index scan remains competitive with the table scan all the way up to about 4M result rows or $2^{-4}$ of the rows in the table.

Despite its advantage over the traditional index scan, its performance is also poor for very large results. If all rows in the table satisfy the query predicate, the performance of the improved index scan is about 2½ times worse than a table scan. While a factor of 2½ is certainly painful, it is less so than the cost of a traditional index scan, which would exceed the cost of a table scan by multiple orders of magnitude.

The obvious promising insight from Figure 1 is that robust execution seems possible; the obvious painful insight is that the improved index scan as implemented in this system is not quite robust enough yet. Putting these insights into a larger perspective, one may hope that a single query execution plan might eventually be superior or at least competitive across the entire range. If that can be achieved, an erroneous choice during compile-time query optimization can be avoided by eliminating the need to choose. On the other hand, this query execution engine has not yet reached the required level of sophistication and robustness. Given the fairly simple techniques we believe are underlying the "improved" plan in Figure 1, we hope that we can devise appropriate run-time techniques and that our new benchmark will guide us there.

Just as Figure 1 immediately enables some observations, insights, and perspective on the entire research effort, other visualizations enable additional insights into additional aspects of robustness. We have found those visualizations helpful for individual operations such as index scans and for plan fragments such as scans of multiple indexes combined by index intersection. Visual images greatly assist in identifying poor scalability or robustness, discontinuities in actual execution costs, etc. In other words, they help in analyzing and reasoning about query execution algorithms, their implementations, entire query execution plans or fragments thereof, and the query execution architecture.

These visualizations can be employed by database software vendors in order to target their improvements in query execution, indexing techniques, and query optimization; or they can be used by database administrators forced to hint specific query execution plans due to unsatisfactory performance or robustness of query execution. The purpose of this paper is to share some of the visualizations we have found particularly helpful as well as some of the insights we have obtained so far.

## 2  Prior work

Depending on the perspective with which they view robustness, some efforts to achieve database system robustness treat query processing internals as a black box. Robust physical design endeavors to make good physical design decisions despite uncertain information about workload characteristics [EA08]. Workload management policies cope with unpredictable queries while still meeting service level objectives, but do not modify the database engine [KKDK07]. Other researchers concentrate on topics related to robust query processing, such as query robustness, database performance benchmarks, and visualization of query plan costs. These efforts enlightened us, but we draw the following distinctions between these prior efforts and our own.

### 2.1  Query optimization

Many researchers focus on the query optimizer's compile-time choices to the exclusion of the executor's run-time performance, and propose methods by which the query optimizer can detect and compensate for errors in cardinality estimation. Systems like COMET or the IBM LEO (LEarning Optimizer) or more recently Microsoft SQL Server use monitoring and feedback to repair incorrect cardinality estimates and statistics [ZHJLZ05, SLMK01, ML02]. Babu, Bizarro, Kabra, Markl, and their co-authors propose different ways to recover from bad cardinality estimates by dynamically re-optimizing or otherwise dynamically changing the query's plan [BBD05, BBDW05, MRSLPC04]. Similarly, Ioannidis et al. propose parametric query optimization methods whereby multiple alternative plans are identified at compile-time, after which an actual plan is selected at run-time when the actual parameter values are known [INSS98]. In contrast, our current focus is on the executor's role – specifically, on measuring the performance of a given plan across a wide range of conditions.

## 2.2 Query execution

A number of other researchers evaluate the run-time performance of implementations of individual operators but do not attempt to capture this information into a form that the query optimizer (or database developers) could use at compile-time to compare how performance degrades as conditions change. For example, Schneider and DeWitt analyze and compare four parallel join algorithms under a variety of conditions, including a comparison of how their performance changes with varying amounts of available memory [SD89]. Gupta et al. compare the performance of property map and bitmap indexing techniques, including a discussion of how parameters such as block size, selectivity, and cardinality impact performance [GDB02]. Cole and Graefe define primitives that enable dynamic plans to be constructed at compile-time while postponing certain decisions until run-time so as to accommodate errors in selectivity estimation, unknown run-time bindings for host variables in embedded queries, and unpredictable availability of resources [CG94, G93, GW89]. However, the focus of such efforts is entirely on the run-time behavior of the operators themselves, as opposed to our own goal of enabling the evaluation and comparison of the robustness characteristics of operator implementations.

## 2.3 Database benchmarking

Most database benchmarks attempt to create a portable embodiment of particular workloads so as to give a standard measure that can be used to assess the relative run-time performance of database systems. For example, the TPC suite of commercial database benchmarks specifies standard workloads and scenarios that can be used to provide relative price/performance comparisons of database systems [OP06, PF00]. Vieira and Madeira propose a dependability benchmark that measures the ability of a database system to recover from various types of system failure [VM03, VM02]. None of these, however, focus attention on the relative performance of a query plan as a function of the actual runtime conditions, as we do.

## 2.4 Plan visualization

We owe a debt to Harista et al. [HDH08, HDH07, RH05, SD04], who have produced an intriguing suite of papers that analyze the compile-time choices of database query optimizers over the relational selectivity space in terms of the area each plan covers as well as the estimated cost of those plans. They visualize the query optimizer's estimated plan costs under a full range of selectivities [HDH08, DBHH08, SD04]. In [HDH08], they explore how to identify "robust" plans, by which they mean a single plan that is within a certain threshold (e.g., 20%) of all the "best" plans across the entire selectivity space. If no single plan meets that threshold, then they would "fail" to find a robust plan and stick with the original plan. That is, they compare different visualizations of compile-time cost estimates in order to identify plans with cost estimates that fall within some threshold of some optimal plan's cost estimate. Our interest, on the other hand, lies in capturing and visualizing how a plan's performance degrades as work increases or resources decrease so that we can reason about this knowledge. In addition, their visualizations present compile-time cost estimates; they do not consider the actual run-time performance of the plans that they visualize. We, on the other hand, present robustness maps based upon run-time performance measurements.

## 3 Plan robustness maps

One problem with robustness in query processing is the number of factors that can affect performance. Prior research indicates that the strongest influences are data volume (both input and output sizes, both record count and record lengths), skew (non-uniform value distributions and duplicate key values), and resources (memory, I/O bandwidth, etc.).

In many cases, visualizing a single factor is sufficient to discover a discontinuous or counter-intuitive effect on performance. In those cases, visualizations with a one-dimensional parameter space are sufficient. For example, in Figure 1, the parameter "output size" is sufficient to illustrate stark differences among the techniques and the value of the improved index scan.

In other cases, multiple parameters interact to create a discontinuous or counter-intuitive effect on performance. The human limit to three-dimensional perception and the one dimension required for performance restrict effective visualizations to two-dimensional parameter spaces. Examples will be discussed shortly.

At this point, we cannot offer specific novel suggestions for the pairs of parameters to investigate. Many of the relevant parameters have already been identified and have found their way into cost functions employed for query optimization. The most promising candidates for analysis are cost function parameters used in complex and conditional expressions. Visualizing the effects of these parameters and their pair-wise interactions is likely to yield concrete and often immediate opportunities for improvement in the query execution algorithms.

In order to prepare a visualization of robustness in query execution, we eliminate choices in query optimization using hints on index usage, join order, join algorithm, and memory allocation. Data and memory sizes were chosen to be realistic, typical, and (to be honest) conveniently available in our research environment.

Other sizes may lead to new insights not available in our experimental setup.

One parameter we have been ignoring in our research to date is the multi-programming level, in particular concurrency of queries and updates but also concurrency of multiple read-only queries, e.g., how shared scans or a large buffer pool affect robustness. In addition, we suspect that robustness of mixed-workload performance could be improved by multi-version concurrency control. If the workload includes utilities that modify logical and physical database design, multi-version concurrency control might be required not only for the data but also for the database catalogs, with a variety of implications for the execution environment, for example, in plan caching and recompilation.

## 3.1 1-dimensional parameter spaces

Figure 1 is an example of the simplest visualization of performance and robustness. One of the first things to verify in such a diagram is that the actual execution cost is monotonic across the parameter space. For example, fetching rows should become more expensive with additional rows; if cases exist in which fetching more rows is cheaper than fetching fewer rows, something is amiss. For example, the governing policy or some implementation mechanisms might be faulty in the algorithms that switch to pre-fetching large pages instead of fetching individual pages as needed. Moreover, the cost curve should flatten, i.e., its first derivative should monotonically decrease. Fetching more rows should cost more, but the difference between fetching 100 and 200 rows should not be greater than between fetching 1,000 and 1,100 rows. This last condition is not true for the improved index scan in Figure 1 as it shows a flat cost growth followed by a steeper cost growth for very large result sizes.

Figure 2 shows the performance of plans for a simple query similar to the query of Figure 1, with two differences. First, performance is shown not in absolute times but relative to the best plan for each point in the parameter space. This type of diagram is most appropriate if the absolute performance varies very widely across the parameter space. In Figure 1, for example, the left-most data point still represents an output size of about 900 rows (60M×$2^{-16}$). Even with a logarithmic scale for query execution costs, extending the diagram all the way to 1 output row would have increased its height or reduced its vertical resolution by a factor of 2½. Illustrating the relative performance of all plans may permit better resolution and better use of the space available for a diagram.

Second, additional query execution plans are included, namely multi-index plans that join non-clustered indexes such that the join result covers the query even if no single non-clustered index does. These index joins are performed by alternative join algorithms and using alternative join orders.

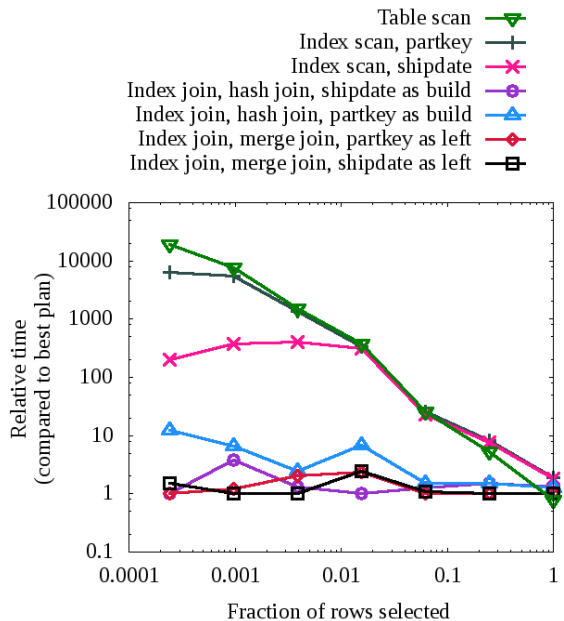

Figure 2. Advanced selection plans.

When comparing query execution plans for a given query, one needs to decide which classes of query execution plans to include: only those actually considered by the system under investigation, those that could be forced by some means or other including alternative syntax (e.g., index intersection by means of multiple query aliases for the same database table), those that could be enabled only by an alternative database design (e.g., two-column indexes), or finally those that could be realized only with additional implementation effort by the software vendor (e.g., bitmap indexes, bitmap-driven sorting or intersection). Obtaining actual execution costs for the last class might require experiments using a competing database system that is more advanced in specific query execution techniques. The most appropriate choice depends on the power one has over design and future improvements of system components. For example, one should consider plans enabled by alternative syntax if one can influence the rewrite capabilities in the query optimization steps.

In addition, one has to decide whether to use linear or logarithmic scales. Logarithmic scales on both axes permit reasonably detailed insight at both ends of the spectrum of possible parameter values. Finally, curves can indicate absolute performance or performance relative to the best plan for any one point in the parameter space, where the definition for "best" might include any of the classes of query execution plans above.

## 3.2 2-dimensional parameter spaces

The limitation to a single dimension within the parameter space both focuses and limits the insights; the interaction of dimensions must also be considered.

The number of possible parameters may be very high – multiple formal query parameters with run-time bindings; resource availability such as memory, processing bandwidth, I/O bandwidth, and interconnection bandwidth; and intermediate result sizes due to predicates (selection, joins), aggregation (projection, duplicate removal), and set operations (intersection, union, difference). Unfortunately, visualization practically forces us to consider two dimensions at-a-time and to rotate through pairs of dimensions.

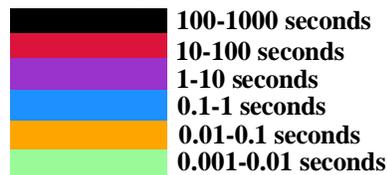

Figure 3. Color code for 2-D maps.

Figure 3 shows the mapping from elapsed times to colors in the following maps, from green to red and finally black (light gray to black in monochrome) with each color difference indicating an order of magnitude.

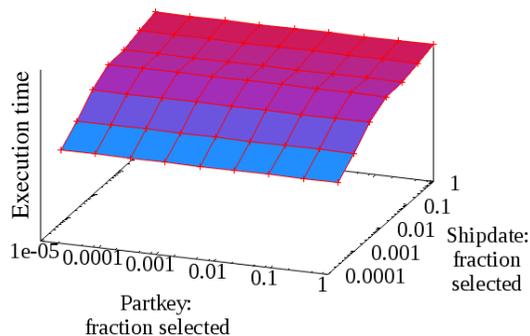

Figure 4. Two-predicate single-index selection.

Figure 4 shows the execution cost for a query restricting two columns of a table. This particular query execution plan scans a single-column index and applies the second predicate only after fetching entire rows from the table's main storage structure. The two dimensions shown are the selectivities of the two predicate clauses. The third dimension is execution time, ranging from 4 seconds to 890 seconds.

Not surprisingly, and immediately visible in the diagram, the two dimensions have very difference effects. In fact, it seems that one of the predicates has practically no effect at all, namely the predicate that can be evaluated only after fetching entire rows. In a way, this diagram is not surprising, which itself is reassuring because index scans perform as expected and as coded in the cost calculations during query optimization. The value of the diagram is its lack of surprise; actual behavior equal to the anticipated behavior (reflected correctly in the cost function used during query optimization) is worth verifying. Figure 4 shows the robust query execution technology from Figure 1. While it is barely visible here, Figure 1 illustrates it very succinctly, demonstrating the value of visualizations using both one-dimensional and two-dimensional parameter spaces.

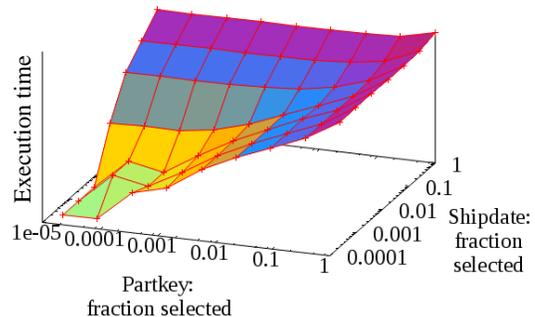

Figure 5. Two-index merge join.

Figure 5 shows the execution cost for an alternative query execution plan, namely scans of two single-column non-clustered indexes combined by a merge join. Other than some measurement flukes in the sub-second range (front left, green), the symmetry in this diagram indicates that the two dimensions have very similar effects. Hash join plans perform better in some cases but do not exhibit this symmetry, as predicted also in our prior research [GLS94]. In fact, our prior research and its presentation would have benefited greatly from visualizations of this type.

## 3.3 Relative performance

In addition to those two plans, we ran five additional alternative query execution plans for this very simple query. These included a no-index table scan (actually, scanning a clustered index organized on an entirely unrelated column), the plan using a single-column non-clustered index for the other predicate clause, and three other plans combining two single-column non-clustered indexes (using merge join or hash join each in two join orders). We then plotted the relative performance of each individual plan compared to the optimal plan at each point in the parameter space. A given plan is optimal if its performance is equal to the optimal performance among all plans, i.e., the quotient of costs is 1. A plan is sub-optimal if the quotient is much higher than 1.

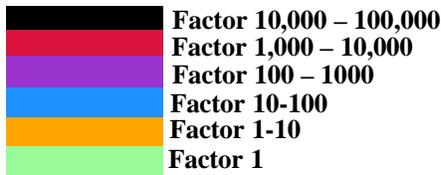

Figure 6. Color code for relative performance.

Figure 6 shows the color code for the following diagrams. One might argue whether a color step for each order of magnitude is appropriate. For example, one might want to focus on small factors such as plans with a cost within a factor of 2 of the cost of the best plan. On the other hand, it seems surprising that a range of five orders of magnitude is required. One cannot but wonder whether consistent and ubiquitous implementation of robust query execution techniques such as those illustrated in Figure 1 would reduce the cost factor of the worst query execution plans or of query execution plans at their worst performance ratio.

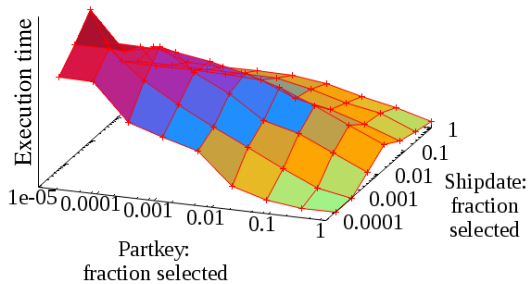

Figure 7. Performance of a single-index scan relative to the best of seven plans.

Figure 7 shows the same data as Figure 4 yet with performance indicated in terms of the relative difference to the best plan at each point. This type of diagram makes it immediately obvious that this plan is optimal only in a small part of the parameter space. Moreover, this region is not continuous, which is rather surprising. Finally, while the absolute performance shown in Figure 4 is fairly smooth, the relative performance shown in Figure 7 is not smooth indicating that the costs of best plans are not smooth. The maximal difference is a factor of 101,000. Thus, while the plan is optimal in some regions of the parameter space, its worst relative performance is so poor that it would likely disrupt data center operation.

In addition to the system used in the earlier diagrams, we also studied two additional systems and their four additional query execution plans for the same set of single-column non-clustered indexes as well as six further plans for two alternative two-column non-clustered indexes. It turned out that a covering two-column index is extremely robust but only if fully exploited using MDAM technology [LJBY95].

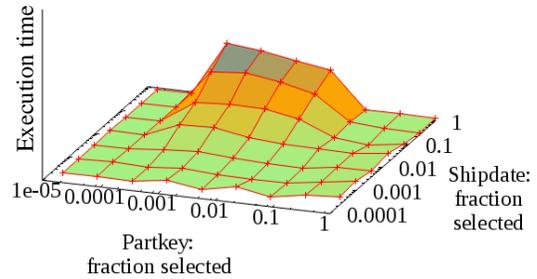

Figure 8. System B relative performance using a two-column index.

Figure 8 shows the relative performance of a plan with a covering two-column index in an another software system. Due to multi-version concurrency control applied only to rows in the main table, this plan requires fetching full rows. In other words, the space overhead of multi-version concurrency control seems to have forced the developers of this system to apply concurrency control only to rows in the main representation of the table. This meant that they had to forgo the advantages of covering non-clustered indexes, including joins of multiple non-clustered indexes.

In this query execution plan, rows to be fetched are sorted very efficiently using a bitmap. This plan is close to optimal in this system over a much larger region of the parameter space. Moreover, its worst quotient is not as bad as the one of the prior plan shown in Figure 7. Thus, if the actual value of parameters is not known at compile-time, this plan is probably much more desirable even if the plans of Figure 4 and Figure 5 are judged more efficient at compile-time based on anticipated predicate selectivities. In other words, robustness might well trump performance in those situations.

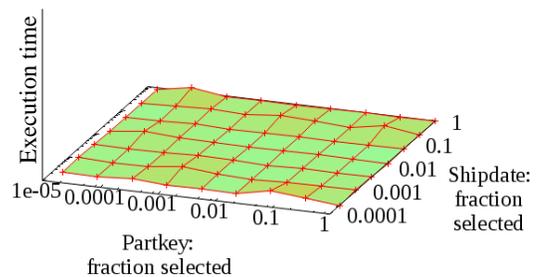

Figure 9. System C relative performance using a two-column index and MDAM algorithms.

Figure 9 shows the most robust plan in a third system on we ran experiments. The relative performance is reasonable across the entire parameter space, albeit not optimal. The foundation of this consistent performance is a very sophisticated scan for multi-column indexes described as multi-dimensional B-tree access in [LJBY95]. Notice that very data points indicate that this plan is the best query execution plan (indicated by a cost factor 1 or a light green color).

Reflecting on the visualization techniques employed here, these diagrams enable rapid verification of expected performance, testing of hypotheses, and insight into absolute and relative performance of alternative query execution plans. Moreover, even for this very simple query, there is a plethora of query execution plans. Investigating many plans over a parameter space with multiple dimensions is possible only with efficient visualizations.

Two opportunities have not been pursued in this paper. First, we have not mapped worst performance, i.e., particularly dangerous plans and the relative performance of plans compared to how bad performance could be. Second, we have not yet compared multiple systems and their available plans. The first system had only 7 plans for this simple two-predicate query; the other two systems had 4 additional plans each for a total of 13 distinct plans across all systems.

Alternative software development activities could ensue after study of these visualizations. First, one can focus on improving the performance of the best plan at some points deemed important within the parameter space – this is the traditional focus on achievable performance. Second, one can focus on the performance of the plan with the broadest region of acceptable performance and then improve its performance in the regions of the parameter space where the plan's performance is poor – this would be a focus on the robustness of a specific plan and, if that plan is chosen during query optimization, on robustness of query processing as a whole.

## 3.4 Mapping regions of optimality

Given the discussions above, one might be tempted to "pull it all together" and show a single map with all possible query execution plans, indicating the best plan for each point and region in the parameter space, perhaps using a color for each plan. The most interesting aspects of these maps would be the size and the shape of each plan's optimality region. Ideally, these regions would be continuous, simple shapes.

For query execution, it might be interesting to focus on irregular shapes of optimality regions – chances are good that some implementation idiosyncrasy rather than the algorithm itself causes the irregular shape. Removal of such idiosyncrasies may lead to more efficient as well as more robust query execution.

In addition, it might be useful to explore techniques that enlarge the largest region, possibly even eliminating some smaller regions and thus some plans from the map of optimality. Every plan eliminated from this map implies that query optimization need not consider this plan and thus cannot err in the decision whether to employ it. Reducing the plan space in query optimization improves query compilation time, and more importantly contributes to the robustness of query optimization.

For query optimization, it might be interesting to explore alternative plans in the order of region sizes. This heuristic might find a good cost bound quickly such that branch-and-bound cost-based pruning can reduce the overall query optimization effort.

In order to reduce the query optimization effort, one might also want to consider fewer plans. This is reasonable if those plans promise both acceptable performance and robustness across the entire parameter space.

Unfortunately, there might be multiple optimal plans (within the measurement error or user tolerance) for any point, thus many points would need to have multiple colors.

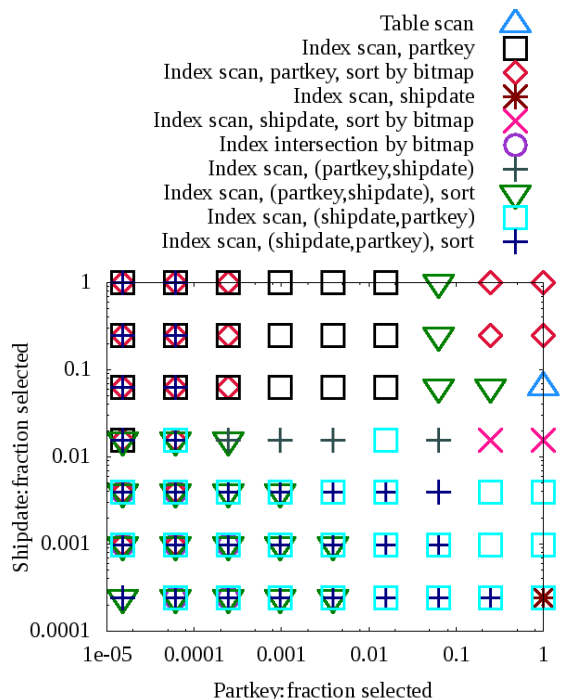

Figure 10. Optimal plans.

Figure 10 illustrates the problem. Most points in the parameter space have multiple optimal plans (within 0.1 sec measurement error). In fact, rather than looking at optimality, one should neglect all small differences. E.g., two plans with actual execution costs within 1% of each other are practically equivalent. Whether this tolerance ends at 1% difference, at 20% difference, or at a factor of 2 depends on one's tradeoff between performance and robustness. Ultimately, this is the tradeoff between the expense of system resources and the expense of human effort for tuning and problem resolution.

Variants of Figure 8 and Figure 9 can be used to show the region of optimality for a specific plan. Since the number of plans that may cover any one point in the parameter space, shading using two colors is not sufficient, but a diagram with points shaded in a large number of colors seems more confusing than illuminating. Thus, this type of diagram inherently requires one diagram per plan and thus many diagrams.

## 4  Summary and conclusions

The run-time performance of any query plan can vary dramatically depending on execution conditions such as actual predicate selectivity and contention for memory and other resources. Such execution conditions vary unpredictably, leading to the unexpectedly long-running queries that plague database users and administrators today. Thus, robust query processing reduces cost of ownership by reducing the need for human intervention.

In general, robustness in database query processing can be improved by modifications in query optimization, query execution, workload management, and other components. Our focus is on query execution. Our immediate work has been on visualizing query execution algorithms and plan fragments in order to understand their behavior across a wide range of unexpected situations.

We find that alternative visualization techniques reveal different insights. We have introduced robustness maps with one- and two-dimensional parameter spaces and have discussed how to interpret them, including a demonstration of how to detect landmarks that appear on these maps and a discussion of their implications for robustness.

Visualizing the performance of specific algorithms, their implementations, and plan fragments using those algorithms shows strengths and weaknesses. We believe that adaptive techniques during run-time query execution can have as great an impact on robust query processing as plan choices during compile-time query optimization. Such adaptive run-time techniques pertain to data volumes, resource availability including memory, and the specifics of the memory hierarchy.

Our immediate next step is to extend this analysis and its visualization to additional query execution algorithms including sort, aggregation, join algorithms, and join order. For example, we expect that some implementations of sorting spill their entire input to disk if the input size exceeds the memory size by merely a single record. Those sort implementations lacking graceful degradation will show discontinuous execution costs. Other resources may introduce similar effect, e.g., a sort input exceeding the size of the CPU cache or the size of flash memory.

This will be followed by visualizations of entire query execution plans including parallel ones. With the experience thus gained, we will then define a benchmark that focuses on robustness of query execution and, more generally, of query processing. This benchmark will identify weaknesses in the algorithms and their implementation, track progress against these weaknesses, and permit daily regression testing in order to protect the progress against accidental regression due to other, seemingly unrelated, software changes. Subsequent research will focus on software techniques that improve robustness and thus benchmark results.

## Acknowledgements

We thank Stavros Harizopoulos for his comments on a draft of this paper, Alex Zhang for discussions regarding performance degradation, and Peter Friedenbach for several very useful scripts.